\documentclass[prd,12pt,tightenlines,preprintnumbers,showpacs,%
superscriptaddress,nofootinbib,amsmath,amssymb]{revtex4}

\begin{document}

\preprint{CECS-PHY-06/26} \preprint{hep-th/0612068}

\title{Higher-dimensional AdS waves and \emph{pp}-waves with
conformally related sources}

\author{Eloy Ay\'on-Beato}\email{ayon-at-cecs.cl}
\affiliation{Centro~de~Estudios~Cient\'{\i}ficos~(CECS),%
~Casilla~1469,~Valdivia,~Chile.}
\affiliation{Departamento~de~F\'{\i}sica,~CINVESTAV--IPN,%
~Apdo.~Postal~14--740,~07000,~M\'exico~D.F.,~M\'exico.}
\author{Mokhtar Hassa\"{\i}ne}\email{hassaine-at-inst.mat.utalca.cl}
\affiliation{Instituto de Matem\'atica y F\'{\i}sica, Universidad de
Talca, Casilla 747, Talca, Chile.}
\affiliation{Centro~de~Estudios~Cient\'{\i}ficos~(CECS),%
~Casilla~1469,~Valdivia,~Chile.}

\begin{abstract}
AdS waves and \emph{pp}-waves can only be supported by pure
radiation fields, for which the only nonvanishing component of the
energy-momentum tensor is the energy density along the retarded
time. We show that the nonminimal coupling of self-gravitating
scalar fields to the higher-dimensional versions of these exact
gravitational waves can be done consistently. In both cases, the
resulting \emph{pure radiation constraints} completely fix the
scalar field dependence and the form of the allowed
self-interactions. More significantly, we establish that the two
sets of pure radiation constraints are conformally related for any
nonminimal coupling, in spite of the fact that the involved
gravitational fields are not necessarily related. In this
correspondence, the potential supporting the AdS waves emerges from
the self-interaction associated to the \emph{pp}-waves and a
self-dual condition naturally satisfied by the \emph{pp}-wave scalar
fields.
\end{abstract}


\maketitle

\section{Introduction}

In the current literature, there exists a wide class of examples for
which the use of conformal symmetries or techniques has been useful
to understand and to solve very specific problems. For example,
conformal transformations of metrics can be used as a mathematical
tool to map the equations of motion of physical systems into
equivalently equations that are more simple to analyze. From the
physical perspective this transformation entails a change of frame,
and in diverse contexts some quantities only acquire a clear
physical interpretation in definite frames. In this spirit, the
derivation of the Bekenstein black hole provided an interesting
illustration \cite{Bekenstein:1974}. Certainly, in this case the
mapping has been operated between a conformal scalar field and a
minimally coupled one, and has permitted to generate a black hole
solution for the conformal theory from the singular solution of the
minimal scalar field theory. It is also natural to ask wether
spacetime metrics which are conformally related may share some
properties, beyond those generated from their common causal
structure for regular conformal factors. For example, in arbitrary
dimension $D$ the Siklos spacetimes \cite{Siklos:1985}, which are
exact gravitational waves traveling along $AdS_D$
\cite{Podolsky:1997ik},\footnote{See
Refs.~\cite{Garcia:1981,Salazar:1983,Garcia:1983,Ozsvath:1985qn} for
the pioneering works studying the propagation of exact gravitational
waves in presence of a cosmological constant, for a review on this
subject see Ref.~\cite{Bicak:1999h}.} can be defined as a conformal
transformation of a \emph{pp}-wave background in the following way
\begin{equation}
ds^2=\frac{l^2}{y^2}\left[-F(u,y,x^i)du^2-2dudv+dy^2
+\delta_{ij}dx^idx^j\right], \label{eq:ppw2AdSw}
\end{equation}
where $i=1,\ldots,D-3$. In this case, the Einstein tensor of the
above AdS wave metric $G_{\alpha\beta}$ and the one associated with
the \emph{pp}-wave metric inside of the square bracket, which we
denote by $\bar{G}_{\alpha\beta}$, satisfy
\begin{equation}
G_{\alpha\beta}+\Lambda{g}_{\alpha\beta}\propto{k}_{\alpha}k_{\beta},
\qquad \bar{G}_{\alpha\beta}\propto{k}_{\alpha}k_{\beta},
\label{eq:property}
\end{equation}
where $k^{\mu}\partial_{\mu}=\partial_v$ is a common null Killing
vector and $\Lambda=-(D-2)(D-1)/(2l^2)$. These properties
(\ref{eq:property}) imply interesting consequences if one considers
the coupling of matter sources to these gravitational waves. Indeed,
in both cases they indicate that the waves can only be supported by
sources behaving as pure radiation fields \cite{Stephani:2003tm},
i.e.\ configurations for which all the components of the
energy-momentum tensor vanish except the energy density along the
retarded time $u$. For a generic matter field this condition imposes
the fulfillment of \emph{pure radiations constraints}, consisting in
demanding the vanishing of all the components of the energy-momentum
tensor except the quoted energy density. These conditions are strong
requirements on the possible choices of matter sources to consider.
Recently, we have been concerned by this problem in three dimensions
as we were interested on the generation of exact gravitational waves
propagating on flat space
\cite{Deser:2004wd,Ayon-Beato:2004fq,Ayon-Beato:2005bm,Jackiw:2005an}
and AdS space \cite{Ayon-Beato:2005qq}, respectively. We have shown
that self-gravitating scalar fields nonminimally coupled to these
gravitational waves do not yield to inconsistencies, instead the
pure radiations constraints have nontrivial solutions in these cases
characterized by various interesting properties due to this
particular coupling \cite{Ayon-Beato:2005qq}. Among these
curiosities, we have shown that the three-dimensional scalar source
supporting an AdS wave \cite{Ayon-Beato:2005qq} and the one
compatible with the \emph{pp}-wave \cite{Ayon-Beato:2005bm} are
conformally related in spite of the fact that their involved
backgrounds are not necessarily related. In other words, this means
that in both situations, the source is independent of the structural
metric function $F$ of Eq.~(\ref{eq:ppw2AdSw}), and in some sense is
only sensitive to the general form of the spacetime metric. At our
opinion, it seems interesting to explore wether the analysis
performed in three dimensions is only due to the simplicity of $2+1$
gravity, or it is generic to any dimension. The main purpose of this
paper is to carry out the correspondence mentioned above to higher
dimensions by establishing a conformal mapping between the pure
radiation constraints determining the scalar sources in each
gravitational wave background.

The organization of the paper is the following. We first consider
the problem of scalar fields nonminimally coupled to a
\emph{pp}-wave background in arbitrary dimension. The resulting pure
radiation constraints are solved in full generality and it is shown
that their integration fixes uniquely the matter source without the
explicit knowledge of the structural metric function. In particular,
there exists a unique self-interaction, depending on a single
coupling constant, allowing the scalar field to act as a source of
the \emph{pp}-wave. In the third section, we reproduce the same
analysis in the context of scalar fields nonminimally coupled to AdS
waves. The pure radiation constraints are again integrable and the
AdS wave scalar source is completely determined. In this case, the
process singles out a unique self-interaction potential depending on
two coupling constants. In the fourth section of the paper, we
establish a conformal correspondence between the pure radiation
constraints of both systems assuming the scalar fields of both
backgrounds are conformally related. In other words, this means that
starting from a \emph{pp}-wave scalar field configuration one is
able to generate the scalar field configuration supporting the AdS
wave. In this correspondence, we emphasize the problem of the
mismatching of the coupling constants of the potentials, problem
which is only specific to higher dimensions. We provide a recipe
which permits to obtain the exact potential supporting the AdS wave,
with its two coupling constants, starting from the \emph{pp}-wave
one with its single coupling constant. In the last section, we
summarize our results and leave some open questions related to this
work. Two Appendixes are included at the end where the detailed
higher-dimensional field equations on each background are explicitly
written.

\section{\label{sec:ppwave}\emph{pp}-waves supported by nonminimally
coupled scalar fields}

In this first part, we are concerned with scalar field nonminimally
coupled to a \emph{pp}-wave background in $D$ dimensions defined by
the line element
\begin{eqnarray}
d\bar{s}^2 & = &
-\bar{F}(u,x^{\hat{i}})du^2-2dudv+\delta_{\hat{i}\hat{j}}dx^{\hat{i}}
dx^{\hat{j}},\nonumber\\
& = & -\bar{F}(u,y,x^i)du^2-2dudv+dy^2+\delta_{ij}dx^{i} dx^{j},
\label{eq:ppwave}
\end{eqnarray}
where the \emph{plane} wave-fronts of the gravitational wave have
coordinates $x^{\hat{i}}=(y,x^i)$ with $i=1,\ldots,D-3$, and its
\emph{parallel} rays are described by the null covariantly constant
field $k^{\mu}\partial_{\mu}=\partial_v$. The action we are
concerned with is given by
\begin{equation}
\bar{S}(\bar{g}_{\alpha\beta},\bar{\Phi})=\int
d^Dx\,\sqrt{-\bar{g}}\left(\frac{1}{2\kappa}\bar{R}
-\frac{1}{2}\bar{g}^{\alpha\beta}\nabla_{\alpha}\bar{\Phi}
\nabla_{\beta}\bar{\Phi}
-\frac{1}{2}\xi\bar{R}\,\bar{\Phi}^2-\bar{U}(\bar{\Phi})\right),
\label{eq:actionppwave}
\end{equation}
where $\xi$ is the parameter characterizing the nonminimal coupling
to gravity of the scalar field $\bar{\Phi}$, whose self-interaction
potential is described by $\bar{U}(\bar{\Phi})$. For later
convenience we have used the convention that all the bared
quantities are those relatives to the \emph{pp}-wave background. The
field equations obtained by varying the action with respect to the
metric and the scalar field read
\begin{equation}
\bar{G}_{\alpha\beta}={\kappa}\bar{T}_{\alpha\beta},
\label{eq:Einsteineqs}
\end{equation}
and
\begin{equation}
\bar{\Box}\bar{\Phi}=\xi \bar{R}\,\bar{\Phi}
+\frac{\mathrm{d}\bar{U}(\bar{\Phi})}{\mathrm{d}\bar{\Phi}},
\label{eq:waveeq}
\end{equation}
respectively, where the energy-momentum tensor is given by
\begin{equation}
\bar{T}_{\alpha\beta}=\nabla_{\alpha}\bar{\Phi}\nabla_{\beta}\bar{\Phi}
-\bar{g}_{\alpha\beta}\left(
\frac{1}{2}\bar{g}^{\mu\nu}\nabla_{\mu}\bar{\Phi}\nabla_{\nu}\bar{\Phi}
+\bar{U}(\bar{\Phi})\right)
+\xi\left(\bar{g}_{\alpha\beta}\bar{\Box}-\bar{\nabla}_{\alpha}
\bar{\nabla}_{\beta} +\bar{G}_{\alpha\beta}\right)\bar{\Phi}^2.
\label{eq:nrg}
\end{equation}
In order to study these configurations we assume that the null
Killing field $k^{\mu}\partial_{\mu}=\partial_v$ is also a symmetry
of the scalar field, i.e.\ $\bar{\Phi}=\bar{\Phi}(u,x^{\hat{i}})$.
The independent Einstein equations on the \emph{pp}-wave background
are given in Appendix \ref{app:Dpp}. As it was stressed in the
introduction, the structure of the Einstein tensor for the geometry
(\ref{eq:ppwave}) is sketched as
\begin{equation}
\bar{G}_{\alpha\beta}\propto{k}_{\alpha}k_{\beta}, \label{eq:ppwprf}
\end{equation}
which in the coordinates adapted to ${k}^{\mu}$ means that the only
nonvanishing component of the Einstein tensor is the one along the
retarded time $\bar{G}_{uu}$. Hence, all the components of the
energy-momentum tensor except $T_{uu}$ must vanish by virtue of the
Einstein equations, and this is interpreted as the scalar field must
behave like a pure radiation field \cite{Stephani:2003tm}. As we
shall see below the integration of the resulting pure radiation
constraints uniquely determines the matter source. Finally, the
remaining independent Einstein equation, i.e.\ the one along the
component $uu$ (see Appendix \ref{app:Dpp}), allows to derive the
structural metric function $\bar{F}$ in the expression
(\ref{eq:ppwave}).

The independent pure radiation constraints are given by the
following combinations, as can be noticed from Appendix
(\ref{app:Dpp}),
\begin{subequations}\label{eq:ppprc}
\begin{eqnarray}
\bar{T}_{u\hat{i}}&=&0,
\label{eq:ppTui}\\
\bar{T}_{\hat{i}\hat{j}}+\delta_{\hat{i}\hat{j}}\bar{T}_{uv}&=&0,
\label{eq:ppTij}\\
\delta^{\hat{i}\hat{j}}\bar{T}_{\hat{i}\hat{j}}+(D-3)\bar{T}_{uv}&=&0.
\label{eq:pptrTij}
\end{eqnarray}
\end{subequations}
As it also occurs in three dimensions \cite{Ayon-Beato:2005bm}, it
is useful to make the following redefinition of the scalar field
\begin{equation}
\bar{\Phi}=\frac{1}{\bar{\sigma}^{2\xi/(1-4\xi)}},
\label{eq:ppPhi2sigma}
\end{equation}
where we are considered all the possible values of the nonminimal
coupling parameter except $\xi=0$ and $\xi=1/4$. These two cases can
also be studied but their analysis is not essential for our main
task. Using Eq.~(\ref{eq:ppEin}) of Appendix \ref{app:Dpp} the first
two equations (\ref{eq:ppTui}) and (\ref{eq:ppTij}) reduce to
\begin{subequations}\label{eq:eredpp}
\begin{eqnarray}
\partial_{u\hat{i}}\bar{\sigma}&=&0,\label{eq:eredpp1}\\
\partial_{\hat{i}\hat{j}}\bar{\sigma}&=&0,\label{eq:eredpp2}
\end{eqnarray}
\end{subequations}
which imply that the general solution is separable in all the
coordinates and is additionally linear in the planar coordinates of
the wave-front
\begin{equation}
\bar{\sigma}=\bar{k}_{\hat{i}}x^{\hat{i}}+\bar{f}(u)
\label{eq:ppsubssigma}
\end{equation}
where the $\bar{k}_{\hat{i}}$ are $D-2$ arbitrary constants and
$\bar{f}$ is an arbitrary function of the retarded time. Hence, the
integration of the first two pure radiation constraints completely
determines the scalar field. Inserting the obtained expression into
the remaining pure radiation constraint (\ref{eq:pptrTij}), the
allowed potential is singled out as
\begin{equation}
\bar{U}(\bar{\Phi})=\frac{2\xi^2\lambda}
{(1-4\xi)^2}\bar{\Phi}^{(1-2\xi)/\xi}, \label{eq:ppU(Phi)}
\end{equation}
where we have defined the coupling constant by
\begin{equation}\label{eq:lambda}
\lambda=\delta^{\hat{i}\hat{j}}\bar{k}_{\hat{i}} \bar{k}_{\hat{j}}.
\end{equation}
The emergence of such potential is interesting by itself for various
reasons. Firstly, for the conformal coupling in $D$ dimensions,
$\xi=\xi_D=(D-2)/[4(D-1)]$, the above potential reduces to the only
potential compatible with the conformal invariance in higher
dimensions $\bar{U}(\bar{\Phi})\propto\bar{\Phi}^{2D/(D-2)}$. We
stress from now that the potential (\ref{eq:ppU(Phi)}) depends on a
single coupling constant $\lambda$ expressed in terms of the $D-2$
integration constants of the scalar field. The solution of the pure
radiation constraints also allows the existence of nontrivial free
massless configurations for $\lambda=0$, as it can be noticed from
Eqs.~(\ref{eq:lambda}) and (\ref{eq:ppsubssigma}). In this case, the
free scalar field depends only and arbitrarily on the retarded time.
It is also interesting to notice that for the particular value of
the nonminimal parameter $\xi=1/2$, the potential
(\ref{eq:ppU(Phi)}) reduces to a positive constant and hence, the
original problem is equivalent to considering the Einstein equations
with a positive effective cosmological constant in presence of a
free nonminimally coupled scalar field.

For later convenience, we record that some of the pure radiation
constraints (\ref{eq:eredpp}) can be reexpressed compactly in terms
of the scalar field and its allowed self-interaction potential
(\ref{eq:ppU(Phi)}) as
\begin{equation}\label{eq:prcconst}
\partial_{\alpha}\left(\frac{\partial_y\bar{\Phi}}
{\sqrt{\bar{U}\left(\bar{\Phi}\right)}}\right)=0,
\end{equation}
which in addition empathizes that the expression between the
parenthesis is a constant. This condition can be interpreted as a
sort of Bogomolnyi self-dual condition for the system.

In sum, we have seen that the integration of the pure radiation
constraints have completely determined the matter source. It is
interesting to stress that this task have been achieved without
using the structural metric function $\bar{F}$. In other words, this
means that the matter source is only sensitive to the form of the
metric and not to the specific structural metric function. As we
shall see below, this property is also present in the case of AdS
waves. Hence, in order to relate the matter sources of both
backgrounds, it is not necessary to derive the structural metric
functions.

\section{\label{sec:AdSwave}AdS waves supported by nonminimally
coupled scalar fields}

In this section, we are concerned with scalar fields nonminimally
coupled to an AdS wave
\begin{equation}
ds^2=\frac{l^2}{y^2}\left[-F(u,y,x^i)du^2-2dudv+dy^2+dx_idx^i\right],
\label{eq:AdSwave}
\end{equation}
where the wave fronts $\{u,v=\mathrm{const.}\}$ are now hyperboloids
with curvature proportional to $-1/l^2$ and coordinates
$x^{\hat{i}}=(y,x^i)$, $i=1,\ldots,D-3$. The field equations are
those arising from the following action
\begin{equation}
S=\int d^Dx\,\sqrt{-g}\left(\frac{1}{2\kappa} (R+2\Lambda)
-\frac{1}{2}g^{\alpha\beta}\nabla_{\alpha}\Phi\nabla_{\beta}\Phi-\frac{1}{2}\xi
R\,\Phi^2-U(\Phi)\right), \label{eq:action}
\end{equation}
where $\Lambda=-(D-2)(D-1)/(2l^2)$ is the negative cosmological
constant, $\xi$ is the nonminimal coupling parameter, and $U(\Phi)$
is the self-interaction potential. The involved field equations are
the Einstein and the nonlinear Klein--Gordon equations
\begin{equation}
G_{\alpha\beta}+{\Lambda}g_{\alpha\beta}={\kappa}T_{\alpha\beta},
\label{eq:Ein}
\end{equation}
\begin{equation}
\Box\Phi=\xi R\,\Phi+\frac{\mathrm{d}U(\Phi)}{\mathrm{d}\Phi},
\label{eq:KG}
\end{equation}
where the corresponding energy-momentum tensor is defined by
\begin{equation}
T_{\alpha\beta}=\nabla_{\alpha}\Phi\nabla_{\beta}\Phi
-g_{\alpha\beta}\left(\frac{1}{2}g^{\mu\nu}\nabla_{\mu}\Phi
\nabla_{\nu}\Phi+U(\Phi)\right)
+\xi\left(g_{\alpha\beta}\Box-\nabla_{\alpha}\nabla_{\beta}
+G_{\alpha\beta}\right)\Phi^2. \label{eq:emt}
\end{equation}
We now see that the strategy used in the case of the \emph{pp}-wave
background can be exactly reproduced here in order to determine the
allowed matter source for an AdS wave background. The clue lies in
the fact that the Einstein tensor for an AdS wave background has the
following structure
\begin{equation}\label{eq:AdSwprf}
G_{\alpha\beta}+{\Lambda}g_{\alpha\beta}\propto{k}_{\alpha}k_{\beta},
\end{equation}
with $k^{\mu}\partial_{\mu}=\partial_v$, which implies that any
self-gravitating source supporting the wave in the presence of the
negative cosmological constant must behave effectively as a pure
radiation field \cite{Stephani:2003tm}. As a consequence, in the
coordinates of metric (\ref{eq:AdSwave}) the only component of the
Einstein equations (\ref{eq:Ein}) with a nonvanishing left hand side
is the $uu$ one as it occurs in the \emph{pp}-wave background. The
other Einstein equations reduce again to pure radiation constraints.

We assume that the null Killing field $k^{\mu}$ is also a symmetry
of the scalar field which in turn implies that the independent field
equations on the AdS wave background reduce to the ones given in
Appendix \ref{app:DAdS}. As before, the pure radiation constraints
are expressed as
\begin{subequations}\label{eq:AdSprc}
\begin{eqnarray}
\label{eq:AdSTui}
T_{u\hat{i}} &=& 0, \\
\label{eq:AdSTij}
T_{\hat{i}\hat{j}}+T_{uv}\delta_{\hat{i}\hat{j}} &=& 0, \\
\label{eq:AdStrTij}
\delta^{\hat{i}\hat{j}}T_{\hat{i}\hat{j}}+(D-3)T_{uv} &=& 0.
\end{eqnarray}
\end{subequations}
We now consider the following redefinition for the scalar field
\begin{equation}\label{eq:AdSPhi2sigma}
\Phi=\frac1{\sigma^{2\xi/(1-4\xi)}},
\end{equation}
which presents the advantage that the pure radiation constraints
(\ref{eq:AdSTui}) and (\ref{eq:AdSTij}), whose explicit form can be
found in the Eq.~(\ref{eq:AdSEin}) of Appendix \ref{app:DAdS}, are
more simple to tackle
\begin{subequations}\label{eq:Ds_yuij}
\begin{eqnarray}
\label{eq:Ds_yu}
\partial_y\left(y\partial_u\sigma\right)  &=& 0, \\
\label{eq:Ds_ui}
\partial_{ui}^2\sigma                     &=& 0, \\
\label{eq:Ds_yy}
\partial_y\left(y^2\partial_y\sigma\right)&=& 0, \\
\label{eq:Ds_yi}
\partial_y\left(y\partial_i\sigma\right)  &=& 0, \\
\label{eq:Ds_ij}
\partial_{ij}^2\sigma                     &=& 0.
\end{eqnarray}
\end{subequations}
The general solution of (\ref{eq:Ds_yuij}) is given by
\begin{equation}\label{eq:AdSsubssigma}
\sigma(u,y,x^i)=\frac{l}y\left[k_yy+k_ix^i+f(u)\right],
\end{equation}
where $k_y$ and $k_i$ are $D-2$ integration constants and $f$ is a
general function of the retarded time. The remaining pure radiation
constraint (\ref{eq:AdStrTij}) is the one that permits to obtain the
allowed potential. After a tedious but straightforward calculation
we conclude that the only self-interaction potential allowed by the
system is given by
\begin{eqnarray}
U(\Phi)&=&\frac{2\xi\Phi^2}{(1-4\xi)^2}
\biggl(\xi\lambda_1\Phi^{(1-4\xi)/\xi}
-8(D-1)\xi(\xi-\xi_D)\lambda_2\Phi^{(1-4\xi)/(2\xi)}\nonumber\\
&&\qquad\qquad\quad{}+\frac{4D(D-1)}{l^2}(\xi-\xi_D)(\xi-\xi_{D+1})
\biggr), \label{eq:AdSU(Phi)}
\end{eqnarray}
where $\xi_D=(D-2)/[4(D-1)]$ is the conformal coupling in dimension
$D$ and the two coupling constants of the potentials are defined by
\begin{equation}\label{eq:Dl12}
\lambda_1={k_y}^2+\delta^{ij}k_ik_j, \qquad \lambda_2=\frac{k_y}{l}.
\end{equation}
Once again, the emergence of such potential is intriguing for
various reasons. In contrast with the allowed potential in the
\emph{pp}-wave situation (\ref{eq:ppU(Phi)}), the AdS wave potential
(\ref{eq:AdSU(Phi)}) depends on two coupling constants
(\ref{eq:Dl12}). This subtlety is not present in three dimensions
\cite{Ayon-Beato:2005qq} since in this case $k_i=0$, and hence the
coupling constants are related, i.e.\
$\lambda_2=\sqrt{\lambda_1}/l$. This remark will be of importance in
the next section where a correspondence between the two
configurations previously analyzed will be presented. It is also
interesting to note that for the conformal value of the nonminimal
coupling parameter, $\xi=\xi_D$, the expression (\ref{eq:AdSU(Phi)})
also reduces to the conformally invariant potential as it occurs in
the \emph{pp}-wave case. In fact, at the vanishing cosmological
constant limit $(l\to\infty)$, we recover the potential permitted by
the \emph{pp}-wave background (\ref{eq:ppU(Phi)}). Finally, we also
mention that this potential is exactly the one arising in the
context of scalar fields nonminimally coupled to special geometries
without inducing backreaction (the static BTZ black hole
\cite{Ayon-Beato:2004ig,Hassaine:2006gz}, flat space
\cite{Ayon-Beato:2005tu,Demir:2006ed}, and the generalized (A)dS
spacetimes \cite{Ayon-Beato:2006c}). All these examples share a
common feature, namely the existence of nontrivial solutions with a
vanishing energy-momentum tensor called stealth configurations.

\section{\label{sec:AdSw/ppw}The Correspondence}

In this section, we establish a correspondence between the two sets
of pure radiations constraints previously studied. The existence of
a map between the involved sources was first noticed in three
dimension \cite{Ayon-Beato:2005qq}. Here, we prove that this
equivalence is not a mere consequence of the apparent simplicity of
$2+1$ gravity and in fact, it can be extended to higher dimensions.
In a more precise set-up, assuming a conformal relation between the
scalar fields generating the two gravitational waves, we first put
in relation the pure radiation constraints that determine the scalar
field solution in both systems. The remaining pure radiation
constraint is the one that fixes the allowed potential of each
system. The relation between these last two constraints is studied
in the second part because of the subtlety due to the mismatching of
the coupling constants of both potentials.

The functional expressions for both scalar fields, on the one hand
Eqs.~(\ref{eq:ppPhi2sigma}) and (\ref{eq:ppsubssigma}), and on the
other hand Eqs.~(\ref{eq:AdSPhi2sigma}) and (\ref{eq:AdSsubssigma}),
suggest to consider a conformal relation between them in the
following manner
\begin{equation}\label{eq:sconf}
\Phi=\left(\frac{l}{y}\right)^s\bar{\Phi},
\end{equation}
where the conformal weight $s$ is not fixed \emph{ab initio}. Using
this relation we intent to write the pure radiation constraints
resulting from an AdS wave (\ref{eq:AdSprc}) in terms of the ones
implied by the existence of a \emph{pp}-wave (\ref{eq:ppprc}).
Running down the components list of the first two sets of pure
radiation constraints (the ones that fix the scalar field
dependence) we obtain the following relations
\begin{subequations}\label{eq:prccorr1}
\begin{eqnarray}
T_{uy}&=&\left(\frac{l}{y}\right)^{2s}\left(\bar{T}_{uy}-
\left[s(1-4\xi)+2\xi\right]
\frac{\partial_{u}\bar{\Phi}^2}{2y}\right),\\
\nonumber\\
T_{ui}&=&\left(\frac{l}{y}\right)^{2s}\bar{T}_{ui},\\
\nonumber\\
T_{yy}+T_{uv}&=&\left(\frac{l}{y}\right)^{2s}
\left[\bar{T}_{yy}+\bar{T}_{uv} - \left[s(1-4\xi)+2\xi\right]y^{s-1}
\partial_y\left(\frac{\bar{\Phi}^2}{y^s}\right)\right],\\
\nonumber\\
T_{yi}&=&\left(\frac{l}{y}\right)^{2s}\left(\bar{T}_{yi}-
\left[s(1-4\xi)+2\xi\right]
\frac{\partial_{i}\bar{\Phi}^2}{2y}\right),\\
\nonumber\\
T_{ij}+\delta_{ij}T_{uv}&=&\left(\frac{l}{y}\right)^{2s}
\left(\bar{T}_{ij}+\delta_{ij}\bar{T}_{uv}\right).
\end{eqnarray}
\end{subequations}

It is clear from these relations that the particular value of the
conformal weight
\begin{equation}\label{eq:weight}
s=-\frac{2\xi}{1-4\xi},
\end{equation}
seems to play a crucial role, but since it remains to connect the
pure radiation constraints (\ref{eq:pptrTij}) and
(\ref{eq:AdStrTij}), we prefer to keep the weight infixed for now.
The possible relation between these two remaining constraints is
more subtle, since this would imply an interrelation among the
potentials (\ref{eq:ppU(Phi)}) and (\ref{eq:AdSU(Phi)}). As it has
been pointed out previously, on the \emph{pp}-wave side there is
only one coupling constant $\lambda$, in contrast with the AdS wave
case where two a priori independent coupling constants $\lambda_1$
and $\lambda_2$ are present. Hence, in order to establish the
correspondence we need to provide a recipe for choosing the two
coupling constants of the AdS wave source starting from the
\emph{pp}-wave one. This problem does not appear in $2+1$ dimensions
where there is only one wave-front coordinate and only one related
integration constant, giving rise to a single coupling constant for
both gravitational wave backgrounds \cite{Ayon-Beato:2005qq}. In
order to compensate this mismatch, our first election is simple and
consists of choosing $\lambda_1$ coinciding with the single coupling
constant of the potential supporting the \emph{pp}-wave. The second
election is inspired by the self-dual condition (\ref{eq:prcconst})
which implies that the quantity
$(\partial_y\bar{\Phi})/\bar{U}(\bar{\Phi})^{1/2}$ is constant for
the \emph{pp}-wave configuration, and hence this allows us to define
the coupling constant $\lambda_2$ as proportional to this constant.
Using the following two definitions for the coupling constants
\begin{subequations}\label{eq:l12tol}
\begin{eqnarray}
\lambda_1&=&\lambda,\\
\lambda_2&=&-\frac1l\sqrt{\frac{\lambda}{2}}
            \frac{\partial_y\bar{\Phi}}
            {\sqrt{\bar{U}\left(\bar{\Phi}\right)}},
\end{eqnarray}
\end{subequations}
we conclude, after a tedious computation, that the remaining pure
radiation constraints are related as follows
\begin{equation}
\delta^{\hat{i}\hat{j}}T_{\hat{i}\hat{j}}+(D-3)T_{uv}
+\frac{l^2}{y^2}\left[U(\Phi)-V_s(\Phi,y)\right]=
\left(\frac{l}{y}\right)^{2s}
\left(\delta^{\hat{i}\hat{j}}\bar{T}_{\hat{i}\hat{j}}
+(D-3)\bar{T}_{uv}\right),
\end{equation}
where the function $V_s$ depends on the scalar field $\Phi$ and
additionally on the wave-front coordinate $y$ by means of
\begin{eqnarray}\label{eq:V_s}
V_s(\Phi,y)&=& \frac{\lambda_1}{\lambda}\bar{U}(\Phi)
\left(\frac{l}{y}\right)^{[s(4\xi-1)-2\xi]/\xi} +
[s+2\xi(D-1)]\frac{\lambda_2}{\sqrt{\lambda}}\Phi\sqrt{2\bar{U}(\Phi)}
\left(\frac{l}{y}\right)^{[s(4\xi-1)-2\xi]/(2\xi)}
\nonumber\\
&&{}+\frac{\left[s^2+4(D-1)\xi{s}+(D-2)(D-1)\xi\right]}{2l^2}\Phi^2,
\end{eqnarray}
where $\bar{U}(\Phi)$ stands for the functional dependence of the
\emph{pp}-wave potential (\ref{eq:ppU(Phi)}) evaluated on the AdS
wave scalar field.

We are now in position to derive some conclusions. Firstly, as it
was previously mentioned it is clear from the relations
(\ref{eq:prccorr1}) that the involved pure radiation constrains of
both gravitational wave backgrounds are conformally related only if
one choose the conformal weight (\ref{eq:weight}). Secondly, for
such weight the function $V_s$ above becomes $y$-independent and
reduces precisely to the functional expression of the scalar
potential supporting the AdS wave (\ref{eq:AdSU(Phi)}). Hence, this
process automatically select the indicated potentials as the only
ones allowing the conformal mapping between the pure radiation
constraints! It is also interesting to note that for the conformal
coupling in $D$ dimensions, $\xi=\xi_D$, the weight
(\ref{eq:weight}) becomes $s=(2-D)/2$ which is precisely the
conformal weight associated to the conformal Klein--Gordon equation
in $D$ dimensions.

In summary, we have shown that the pure radiation constraints on a
\emph{pp}-wave and an AdS wave are conformally related if one
suppose a conformal relation (\ref{eq:sconf}) between the involved
scalar fields with a conformal weight (\ref{eq:weight}), and
additionally the respective potential (\ref{eq:ppU(Phi)}) and
(\ref{eq:AdSU(Phi)}) are taken in each side using the definitions
(\ref{eq:l12tol}) for the coupling constants, i.e.\
\begin{subequations}
\begin{eqnarray}
T_{u\hat{i}}&=&\left(\frac{l}{y}\right)^{-4\xi/(1-4\xi)}
\bar{T}_{u\hat{i}},\\
T_{\hat{i}\hat{j}}+\delta_{\hat{i}\hat{j}}T_{uv}&=&
\left(\frac{l}{y}\right)^{-4\xi/(1-4\xi)}
\left(\bar{T}_{\hat{i}\hat{j}}+\delta_{\hat{i}\hat{j}}\bar{T}_{uv}
\right),\\
\delta^{\hat{i}\hat{j}}T_{\hat{i}\hat{j}}+(D-3)T_{uv} &=&
\left(\frac{l}{y}\right)^{-4\xi/(1-4\xi)}
\left(\delta^{\hat{i}\hat{j}}\bar{T}_{\hat{i}\hat{j}}
+(D-3)\bar{T}_{uv}\right).
\end{eqnarray}
\end{subequations}

A similar conclusion can be achieved by studying the wave equation
for the scalar field. Using the conformal weight (\ref{eq:weight})
and the relations (\ref{eq:l12tol}) between the coupling constants
of both potentials, a conformal relation between the Klein--Gordon
equations can be also achieved for any generic nonminimal coupling
$\xi$,
\begin{equation}
\Box\Phi-\xi{R}\,\Phi-\frac{\mathrm{d}U(\Phi)}{\mathrm{d}\Phi}
=\left(\frac{l}{y}\right)^{2(2\xi-1)/(1-4\xi)}
\left(\bar{\Box}\bar{\Phi} -\xi\bar{R}\,\bar{\Phi}
-\frac{\mathrm{d}\bar{U}(\bar{\Phi})}{\mathrm{d}\bar{\Phi}}\right).
\end{equation}
This relation is far from obvious since it usually only works in the
case of the conformal coupling and taking in both sides of the
equation the unique potential that does not spoil the conformal
invariance. A fact which is straightforwardly recovered in the above
expression just taking $\xi=\xi_D$.

\section{\label{sec:conclu}Conclusions}

Here, we have been concerned with the \emph{pp}-wave and the AdS
wave backgrounds in arbitrary dimension. These spacetimes share in
common that their coupling to a matter source is accompanied by a
strong restriction, namely the source field must behave like a pure
radiation field. The elaboration of this work through two concise
examples in arbitrary dimensions has opened a number of questions
that we would like to comment.

In this paper, we have shown that the nonminimal coupling of scalar
fields to these particular spacetimes can be realized consistently,
and the most general scalar field configurations consistent with the
only symmetry of the problem have been derived. In this first result
there is an interesting contrast between the strong restriction
imposed by the spacetimes and the fact that the most general
solution of an higher-dimensional problem with only one symmetry can
be obtained. Moreover, it is obvious form our study that not any
matter field can act as a source for these backgrounds. In view of
this, it is legitimate to go into thoroughly and ask what are the
characteristics that a matter action must possess in order to couple
consistently with these peculiar spacetimes. For example, it is
clear that since the \emph{pp}-wave and the AdS wave metrics possess
a null Killing field together with the fact that their Einstein
tensors have the structure (\ref{eq:property}), automatically impose
an on-shell traceless condition on the energy-momentum tensor of the
matter source.

In the analysis of the pure radiation constraints, we have put in
evidence the analogies existing between both backgrounds. Indeed, in
each case, the same combinations of the energy-momentum tensor
components give rise to the independent pure radiation constraints.
Moreover, these combinations uniquely fix the scalar field source,
that means not only the local expression for the scalar field but
also the unique self-interactions allowing the existence of the
whole configuration. Furthermore, this derivation has been done
without the explicit knowledge of the structural metric function,
suggesting that the source is only sensitive to the general form of
the metric. This property by itself is very intriguing and unusual
in gravitational physics due to the strongly coupled behavior
inherent to the matter/gravity interaction; as it is well known
matter acts as the source of spacetime curvature generating the
gravitational potential, but at the same time the spacetime geometry
is the arena where matter fields evolve, i.e.\ matter fields feel
the fingerprints of the gravitational field via its equation of
motion. In the present cases the metric structural functions do not
participate in the Klein--Gordon equations. It is natural to ask
first wether there exist other examples of such behavior in the
current literature. To our knowledge the only similar examples occur
for the so-called stealth configurations for which both matter and
gravity are completely decoupled
\cite{Ayon-Beato:2004ig,Hassaine:2006gz,Ayon-Beato:2005tu,
Ayon-Beato:2006c,Hassaine:2005xg}. The analogies with the stealth
configurations also concern the allowed potentials as it has been
stressed in the present work. For all these reasons, it would be
desirable to have a better understanding of these curiosities from a
mathematical as well as physical point of view.

As said before, the pure radiation constraints impose and single out
a unique form of the potential for each background. In the
\emph{pp}-wave case, the selected self-interaction depends on a
single coupling constant and follows a power-law dependence on the
scalar field. In the AdS wave case, two coupling constants emerge
from the integration of the pure radiation constraints, each one
associated to a different power-law term in the potential,
additionally a third contribution also appears consisting in a mass
term whose mass scale is fixed by the AdS radius. In spite of being
different potentials, in the case of the conformal coupling in $D$
dimensions, both potentials reduce to the conformally invariant
potential. It is appealing that as the nonminimal coupling parameter
takes the conformal value, the allowed potentials precisely reduce
to the conformally invariant one in $D$ dimensions. This may be
think as if for an arbitrary value of the nonminimal coupling
parameter, the system would enjoy a symmetry higher than the
conformal symmetry and reduces to this later as the nonminimal
coupling parameter becomes the conformal one. The conformal
relations established in the previous section brings evidence in
favor of this view. An interesting work will then consist of
studying the dynamical symmetries of the models we have considered
in order to confirm the existence or not of a higher symmetry
including the conformal one.

In the last part, we have extended the analogies observed between
the AdS wave and the \emph{pp}-wave sources by establishing a
conformal correspondence between the pure radiation constraints of
each system. In some sense, this correspondence permits to derive
the scalar field configuration of the AdS wave background from the
\emph{pp}-wave one in a nontrivial way. In this correspondence, the
scalar fields are conformally related with a weight expressed in
terms of the nonminimal coupling parameter independently of the
precise dimension. For the conformal value of the nonminimal
coupling, this weight precisely becomes the conformal weight
associated to the conformal Klein--Gordon equation. The pure
radiation constraints fixing the scalar field dependence are easily
put in equivalence in contrast with the radiation constraints that
single out the self-interactions. Indeed, in this last case, there
is a mismatching between the coupling constants of the respective
potentials. The additional coupling constant in the AdS wave
potential has been shown to be associated to a constant arising from
a self-dual condition naturally satisfied by the \emph{pp}-wave
scalar fields. It is far from obvious that the self-gravitating
matter sources generating each backgrounds are in correspondence
even if these backgrounds can be viewed as conformally related. One
may think that the correspondence established here is a sort of
residual conformal symmetry that has its origin on the on-shell
traceless condition of the energy-momentum tensor, a property
usually associated to the conformal invariance of the source. Once
again, it would be of interest to understand the mathematical
structures that are behind of the examples treated in this work.

\begin{acknowledgments}
We thank Alberto Garc\'{\i}a for useful discussions. This work is
partially supported by grants 3020032, 1040921, 7040190, 1051064,
1051084, and 1060831 from FONDECYT, and grants CO1-41639 and
CO2-44598 from CONACyT. Institutional support to the Centro de
Estudios Cient\'{\i}ficos (CECS) from Empresas CMPC is gratefully
acknowledged. CECS is a Millennium Science Institute and is funded
in part by grants from Fundaci\'{o}n Andes and the Tinker
Foundation.
\end{acknowledgments}

\appendix

\section{\label{app:Dpp}field equations for higher dimensional
\emph{pp}-waves}

The independent Einstein equations (\ref{eq:Ein}) for the
energy-momentum (\ref{eq:emt}) on the background of a
$D$-dimensional \emph{pp}-wave (\ref{eq:ppwave}) are given by the
following combination
\begin{eqnarray}
0&=&\bar{G}_{\alpha\beta}-{\kappa}\bar{T}_{\alpha\beta}
+\bar{g}_{\alpha\beta}(\bar{G}_{uv}-{\kappa}\bar{T}_{uv})
\nonumber\\
&=&\left[\frac12(1-\kappa\xi\bar{\Phi}^2)\hat{\triangle}{\bar{F}}
+\kappa\xi\left(\partial_{uu}^2\bar{\Phi}^2
-\frac12\delta^{\hat{i}\hat{j}}
\partial_{\hat{i}}\bar{F}\partial_{\hat{j}}\bar{\Phi}^2\right)
-\kappa(\partial_u\bar{\Phi})^2\right]\delta_\alpha^u\delta_\beta^u
\nonumber\\
&&{}-2\kappa\left(\partial_u\bar{\Phi}\partial_{\hat{i}}\bar{\Phi}
-\xi\partial_{u\hat{i}}^2\bar{\Phi}^2\right)
\delta_{(\alpha}^u\delta_{\beta)}^{\hat{i}}
-\kappa\left(\partial_{\hat{i}}\bar{\Phi}\partial_{\hat{j}}\bar{\Phi}
-\xi\partial_{\hat{i}\hat{j}}^2\bar{\Phi}^2\right)
\delta_\alpha^{\hat{i}}\delta_\beta^{\hat{j}},\label{eq:ppEin}
\end{eqnarray}
and the trace
\begin{equation}\label{eq:pptrace}
0=\bar{g}^{\hat{i}\hat{j}}(\bar{G}_{\hat{i}\hat{j}}
-{\kappa}\bar{T}_{\hat{i}\hat{j}})
+(D-3)(\bar{G}_{uv}-{\kappa}\bar{T}_{uv})
=\kappa\left(\bar{U}(\bar{\Phi})-\frac12\delta^{\hat{i}\hat{j}}
\partial_{\hat{i}}\bar{\Phi}\partial_{\hat{j}}\bar{\Phi}\right),
\end{equation}
where $x^{\hat{i}}=(y,x^i)$, $i=1,\ldots,D-3$, and
$\hat{\triangle}=\delta^{\hat{i}\hat{j}}\partial_{\hat{i}}
\partial_{\hat{j}}$.
It is straightforward to check that the above equations reduce to
its $2+1$ dimensional versions solved in
Ref.~\cite{Ayon-Beato:2005bm}.

\section{\label{app:DAdS}field equations for higher dimensional
AdS waves}

All the information following from Einstein equations (\ref{eq:Ein})
with energy-momentum (\ref{eq:emt}) on the background of a
$D$-dimensional AdS wave (\ref{eq:AdSwave}) is encoded in the
following combination
\begin{eqnarray}
0&=&G_{\alpha\beta}+{\Lambda}g_{\alpha\beta}-{\kappa}T_{\alpha\beta}
+\frac{y^2}{l^2}g_{\alpha\beta}(G_{uv}+{\Lambda}g_{uv}-{\kappa}T_{uv})
\nonumber\\
&=&\left[\frac12\frac{l^2}{y^2}(1-\kappa\xi\Phi^2)\Box{F}
+\kappa\xi\left(\partial_{uu}^2\Phi^2
-\frac12\delta^{\hat{i}\hat{j}}
\partial_{\hat{i}}F\partial_{\hat{j}}\Phi^2\right)
-\kappa(\partial_u\Phi)^2\right]\delta_\alpha^u\delta_\beta^u
\nonumber\\
&&{}-2\kappa\left(\partial_u\Phi\partial_y\Phi
-\frac{\xi}y\partial_y\left(y\partial_u\Phi^2\right)\right)
\delta_{(\alpha}^u\delta_{\beta)}^y
-2\kappa\left(\partial_u\Phi\partial_i\Phi
-\xi\partial_{ui}^2\Phi^2\right)\delta_{(\alpha}^u\delta_{\beta)}^i
\nonumber\\
&&{}-\kappa\left((\partial_y\Phi)^2
-\frac{\xi}{y^2}\partial_y\left(y^2\partial_y\Phi^2\right)\right)
\delta_\alpha^y\delta_\beta^y
-2\kappa\left(\partial_y\Phi\partial_i\Phi
-\frac{\xi}y\partial_y\left(y\partial_i\Phi^2\right)\right)
\delta_{(\alpha}^y\delta_{\beta)}^i
\nonumber\\
&&{}-\kappa\left(\partial_i\Phi\partial_j\Phi
-\xi\partial_{ij}^2\Phi^2\right)\delta_\alpha^i\delta_\beta^j,
\label{eq:AdSEin}
\end{eqnarray}
and the trace
\begin{eqnarray}
0&=&g^{\hat{i}\hat{j}}(G_{\hat{i}\hat{j}}+{\Lambda}g_{\hat{i}\hat{j}}
-{\kappa}T_{\hat{i}\hat{j}})
+(D-3)\frac{y^2}{l^2}(G_{uv}+{\Lambda}g_{uv}-{\kappa}T_{uv})
\nonumber\\
&=&\kappa\left(U(\Phi)+\xi\Lambda\Phi^2
+\xi(D-1)\frac{y}{l^2}\partial_y\Phi^2
-\frac12\frac{y^2}{l^2}\delta^{\hat{i}\hat{j}}
\partial_{\hat{i}}\Phi\partial_{\hat{j}}\Phi\right),\label{eq:AdStrace}
\end{eqnarray}
where $x^{\hat{i}}=(y,x^i)$, $i=1,\ldots,D-3$,
$\Lambda=-(D-2)(D-1)/(2l^2)$, and
\begin{equation}\label{eq:Box}
\Box{F}=\frac{y^2}{l^2}\left[
y^{D-2}\partial_y\left(\frac1{y^{D-2}}\partial_yF\right)
+\triangle{F}\right],
\end{equation}
with $\triangle=\delta^{ij}\partial_i\partial_j$. As before the
above equations becomes the $2+1$ dimensional ones studied in
Ref.~\cite{Ayon-Beato:2005qq}.



\begin{thebibliography}{99}

\bibitem{Bekenstein:1974}
  J.~D.~Bekenstein,
  Annals Phys.\ \textbf{82}, 535 (1974);
  \textbf{91}, 75 (1975).

\bibitem{Siklos:1985}
  S.T.C.~Siklos,
  in: \emph{Galaxies, axisymmetric systems and relativity},
  ed. M.A.H. MacCallum (Cambridge Univ. Press, Cambridge 1985).

\bibitem{Podolsky:1997ik}
  J.~Podolsky,
  Class.\ Quant.\ Grav.\ \textbf{15}, 719 (1998)
  [arXiv:gr-qc/9801052].

\bibitem{Garcia:1981}
  A.~Garc\'{\i}a and J.~Pleba\'nski,
  J. Math. Phys. \textbf{22}, 2655 (1981).

\bibitem{Salazar:1983}
  H. Salazar, A.~Garc\'{\i}a, and J.~Pleba\'nski,
  J. Math. Phys. \textbf{24}, 2191 (1983).

\bibitem{Garcia:1983}
  A.~Garc\'{\i}a, Nuovo Cim. B \textbf{78}, 255 (1983).

\bibitem{Ozsvath:1985qn}
  I.~Ozsvath, I.~Robinson, and K.~Rozga,
  J.\ Math.\ Phys.\ \textbf{26}, 1755 (1985).

\bibitem{Bicak:1999h}
  J.~Bicak and J.~Podolsky,
  J.\ Math.\ Phys.\ \textbf{40}, 4495 (1999)
  [arXiv:gr-qc/9907048];
  4506 (1999)
  [arXiv:gr-qc/9907049].

\bibitem{Stephani:2003tm}
  H.~Stephani, D.~Kramer, M.~MacCallum, C.~Hoenselaers, and E.~Herlt,
  \emph{Exact solutions of Einstein's field equations}
  (Cambridge University Press, Cambridge 2003).

\bibitem{Deser:2004wd}
  S.~Deser, R.~Jackiw, and S.~Y.~Pi,
  Acta Phys.\ Polon.\ B \textbf{36}, 27 (2005)
  [arXiv:gr-qc/0409011].

\bibitem{Ayon-Beato:2004fq}
  E.~Ay\'{o}n--Beato and M.~Hassa\"{\i}ne,
  Annals Phys. \textbf{317}, 175 (2005)
  [arXiv:hep-th/0409150].

\bibitem{Ayon-Beato:2005bm}
  E.~Ay\'{o}n--Beato and M.~Hassa\"{\i}ne,
  Phys.\ Rev.\ D \textbf{71}, 084004 (2005)
  [arXiv:hep-th/0501040].

\bibitem{Jackiw:2005an}
  R.~Jackiw,
  AIP Conf.\ Proc.\ \textbf{805}, 283 (2006)
  [arXiv:gr-qc/0509035].

\bibitem{Ayon-Beato:2005qq}
  E.~Ay\'{o}n--Beato and M.~Hassa\"{\i}ne,
  Phys.\ Rev.\ D \textbf{73}, 104001 (2006)
  [arXiv:hep-th/0512074].

\bibitem{Ayon-Beato:2004ig}
  E.~Ay\'{o}n--Beato, C.~Mart\'{\i}nez, and J.~Zanelli,
  Gen.\ Rel.\ Grav.\ \textbf{38}, 145 (2006)
  [arXiv:hep-th/0403228].

\bibitem{Hassaine:2006gz}
  M.~Hassa\"{\i}ne,
  J.\ Phys.\ A \textbf{39}, 8675 (2006)
  [arXiv:hep-th/0606159].

\bibitem{Ayon-Beato:2005tu}
  E.~Ay\'{o}n--Beato, C.~Mart\'{\i}nez, R.~Troncoso, and J.~Zanelli,
  Phys.\ Rev.\ D \textbf{71}, 104037 (2005)
  [arXiv:hep-th/0505086].

\bibitem{Demir:2006ed}
  D.~A.~Demir and B.~Pulice,
  Phys.\ Lett.\ B \textbf{638}, 1 (2006)
  [arXiv:hep-th/0605071].

\bibitem{Ayon-Beato:2006c}
E.~Ay\'{o}n--Beato, C.~Mart\'{\i}nez, R.~Troncoso, and J.~Zanelli,
``Stealths on AdS,'' in preparation.

\bibitem{Hassaine:2005xg}
  M.~Hassaine,
  J.\ Math.\ Phys.\  {\bf 47}, 033101 (2006)
  [arXiv:hep-th/0511243].

\end{thebibliography}
\end{document}